# Evidences of vortex curvature and anisotropic pinning in superconducting films by quantitative magneto-optics

F. Laviano, D. Botta, A. Chiodoni, R. Gerbaldo, G. Ghigo, L. Gozzelino, and E. Mezzetti

Department of Physics, Politecnico di Torino, C.so Duca degli Abruzzi 24, 10129 Torino, Italy.

Istituto Nazionale di Fisica Nucleare, Sez. Torino, via P. Giuria 1, 10125 Torino, Italy.

Istituto Nazionale per la Fisica della Materia, U.d.R. Torino-Politecnico, C.so Duca Degli Abruzzi 24, 10129 Torino, Italy.

Abstract

We present the experimental observation of magnetic field line curvature at the surface of a superconducting film by local quantitative magneto-optics. In addition to the knowledge of the full induction field at the superconductor surface yielding the quantitative observation of the flux line curvature, our analysis method allows also local value measurements of the electrical current density inside the sample. Thus, we study the interplay between the electrodynamic constraints dictated by the film geometry and the pinning properties of the superconductor. In particular, we investigate the anisotropic vortex-pinning, due to columnar defects introduced by heavy ion irradiation, as revealed in the local current density dependence on the vortex curvature during magnetic flux diffusion inside the superconducting film.

# 1. Introduction

Superconducting films, with transport current or applied magnetic field perpendicular to the film plane (transverse geometry), present a quite complicated critical state problem [1,2] with respect to other configurations, e.g. long superconductors in parallel applied fields (longitudinal geometry). As theoretically found [3], the critical state problem for thin superconductor, i.e. with thickness $d$ much smaller than lateral extension, is actually three dimensional (3D). For the transverse geometry, the supercurrents flow in the film plane to screen the perpendicular external field, but they induce also a magnetic field parallel to such plane. The *in-plane* magnetic induction components reach the maximum values at the superconductor surface and cancel at the central plane, resulting in strongly tilted magnetic field lines inside and outside the specimen [4]. This magnetic field line curvature has important consequences during vortex diffusion in superconducting films, until the flux lines straighten along the external field direction far above the full penetration field, i.e. when it is commonly stated that demagnetising effects are negligible.

First of all, the Lorenz forces acting on vortices don't lay in the film plane, but they have not zero perpendicular components. Thus, when defects have axial symmetry with respect to the film normal direction, hereafter named $z$ axis, the local critical current is not homogeneously distributed along the film thickness, i.e. the thickness averaged current density depends on the angle $\theta$ [3], which represents the magnetic field line tilt angle away from z axis at the superconductor surface.

Moreover, $\theta$ is not constant, but changes with position inside the sample. Therefore, to account for the flux line curvature, a *local and parallel* measurement process is needed, i.e. we have to know the distributions of all the magnetic induction components. From this point of view, characterizing the electrodynamics of superconductors in transverse geometry is a hard challenge to fulfil in a detailed quantitative fashion.

The influence of flux line curvature on anisotropic pinning sites was indeed recently recognized in superconducting single crystals with columnar defects, where angle resolved magnetization measurements reveal peaks when the applied field direction is slightly misaligned from the tracks [5]. In a pioneering work, Schuster et al. qualitatively demonstrated, by means of the magneto-optical technique [6], how inclined columnar defects with respect to the z axis present, in transverse geometry, were a more efficient pinning than parallel columns.

The magneto optical technique is one of the best candidates for the local and parallel observation of the magnetic moment distribution over the surface of a superconductors. In principle, by the numerical inversion of the Biot-Savart law, it is possible to reconstruct the current density distribution throughout the magnetic induction ($z$ component) measured by magneto-optics in a model independent way [7], but the knowledge of the in-plane magnetic field components is required both for the flux line curvature study and for the quantitative magneto optical (QMO) analysis itself, because the in-plane fields couples with the ferromagnetic indicator films [8].

We extended the correction procedure already elaborated for strip geometry [9] to a model-independent QMO analysis accounting for the in-plane magnetic fields at the superconductor surface [10]. Yielding the local measurement of the actual magnetic induction in the z direction, $B_z(x,y)$, and the evaluation of the in-plane magnetic field components and of current density distribution, our method enables to study the 3D critical state of thin superconductors.

In this paper, we present such QMO analysis about $YBa_2Cu_3O_{7-x}$ (YBCO) films, aimed at studying the interplay among magnetic field lines curvature, intrinsic pinning and extrinsic anisotropic defects correlated with the *z* axis (irradiation induced columns). In particular, we show how our QMO analysis is able to measure the local magnetic field line curvature (section 3) and how it reveals the anisotropic pinning properties of both cases, intrinsic and extrinsic defects, into marked and characteristic dependencies of the local current density on tilt angle θ (section 4). We also discuss the implications about the critical current determination for superconductors in transverse geometry and the effect of thermal fluctuations on the studied system (section 5).

## 2. Experimental details and QMO analysis method

The superconducting samples are $YBa_2Cu_3O_{7-x}$ films grown by thermal co-evaporation on yttria stabilised zirconia (YSZ) substrate with a 40 nm thick $CeO_2$ buffer layer [11]. The YBCO film thickness is 400 nm, its $T_C$ is about 88K and $\Delta T_C$ ~0.7K. After the deposition, the original film, square with 1 cm long sides, was chemically etched in order to obtain smaller squares (in number of nine), with side of about 1.25 mm.

The QMO analysis was performed on the whole set of twin samples, which resulted in excellent reproducibility of the measurement process and in homogeneous superconducting properties between the squared specimens. Here, we present the QMO results on two samples: the squared film (SQ1), showing the least disrupted current flow pattern by macroscopic defects, was chosen to illustrate the flux line curvature imposed by the transverse geometry [12] (optical picture in figure 1(a)), and another squared sample (SQ2), figure 1(b), which was measured before and after the irradiation with $Au^{17+}$ ions at 250 MeV, perpendicular to the film plane and with dose equivalent field $B_\Phi$ of 0.1 T.

The TEM analysis, on remaining twin samples, reveals these films have *c* axis normal to the substrate surface, but their in-plane grain orientation is disordered, as shown in figure 2(a). In figure 2(b), the high resolution cross-section shows the $YBCO/CeO_2$/substrate interfaces; in particular, it is visible a small rounded amorphous region at the superconductor/buffer layer interface. The planar view of figure 2(c) displays the damage produced by the heavy ions crossing the film. We note that extrinsic defects are insulating columns with diameter of few nanometers and they are surrounded by an amorphous zone caused by the heavy ion collision too, as well visible in figure 2(c). These morphologic features give to irradiation induced columnar pins optimal pinning on vortices in superconductors with short superconducting coherence length, such as the high temperature YBCO superconductor [13].

The QMO measurements consisted of several zero field cooling of the superconductor, down to different temperatures, followed by applying a set of increasing magnetic field directed along the z axis. Each applied field was kept constant for 3 s, until a magneto-optical image was acquired by our standard set-up [10]. After the calibration process to convert the measured light intensity to $B_z$ values, the numerical method returns, from each image, the distributions of the full induction field at the superconductor surface ($B_x(x,y,d/2)$, $B_y(x,y,d/2)$ and $B_z(x,y,d/2)$, where the reference $z=0$ plane is at half thickness of the sample), and the local electrical current density averaged over the sample thickness, $J_x(x,y)$ and $J_y(x,y)$. Our method is based on the iterative correction of the pristine calibrated data because of the in-plane magnetic fields induced by the superconducting sample [9]. The iterative algorithm evaluates the in-plane magnetic field components from the current density obtained by Biot-Savart inversion of the $B_z$ data, afterward it corrects the measured $B_z$ data itself until the convergence is reached; for details see [10]. This correction is needed for the quantitative evaluation of the magnetic induction distribution, but it is even weightier for current density reconstruction. In summary, the precise measurement of the out-of-plane component of the magnetic induction, $B_z(x,y,d/2)$, and the evaluation of the in-plane components, enables us to estimate the local tilt angle $\theta(x,y,d/2)$ of the magnetic field lines at the superconductor surface (hereafter the fixed z position at the film surface, d/2, will be omitted), as simply as:

$$\theta(x,y) = \arctan \frac{B_{xy}(x,y)}{B_z(x,y)}.$$

### 3. Magnetic field line curvature

The remarkable demagnetising effects of a superconductor in transverse geometry are well appreciable before the full magnetic flux penetration. A representative state during the magnetic flux diffusion, in the virgin magnetization process of the SQ1 YBCO film, is reported in figure 3, where the $B_z(x,y)$ distribution and the corresponding $J(x,y)$ stream lines are shown. The partly penetrated state is displayed by the magnetic field gradient, generated by the vortices diffused from the edges and arranged in pinning centers, and by the dark area of cushion-like shape, which represents the flux-free region inside the superconductor, where the $B_z$ component is zero.

Because of the transverse geometry, the flux-free volume has a 3D lens-shaped profile [14] and the Meissner currents, as visible by the current stream lines of figure 3, occupy most of this region flowing around it. In fact, the critical state for thin superconductors in transverse geometry develops mainly across the thickness [3] and we expect the magnetic field lines are curved as they hug the internal flux-free zone.

The corresponding 3D-distribution of the magnetic induction measured at the superconductor surface, presented in figure 4, indeed shows all the general features of the transverse geometry demagnetising behaviour. In figure 4(a), the 3D vector plot of B(x,y) demonstrates how the magnetic field generated by the superconductor, the *self-field*, changes both in intensity and in direction with respect to the external field.

Outside the superconducting film, the stray field is superimposed to the external field, so, near the sample edges, the $B_z$ component is increased with respect to the applied magnetic field; the maximum of $B_z$ is located at the center of each edge, whereas the minimum occurs in correspondence of the corners, as also visible in figure 3. Over the specimen, the magnetic field lines changes their direction from the film normal toward the film plane, i.e. they are curved toward the flux-free region inside the sample, as best shown by the projection of B on the *xy* plane, in figure 4(b). Also the finite lateral shape plays an important role as clearly revealed in figure 3 and figure 4, where the characteristic discontinuity lines show up [15]: the in-plane magnetic field changes direction at some of these lines, commonly called **d**+ lines, as the current density does.

As seen by these pictures, the local change of the magnetic field, both in intensity and in direction inside superconductors of finite sizes, makes a local quantitative analysis needed. The quantitative information, provided by the QMO method, can be synthesized through linear profiles of the bi-dimensional (2D) distributions. The $B_z(x,y_0)$ values, i.e. at fixed y and several x, and the in-plane components modulus $|B_{xy}(x,y_0)|$, are plotted in figure 5(a) and 5(b), respectively, in order to show the dynamical evolution of the magnetic field distribution over the SQ1 sample at different applied field. The shape of these profile displays the characteristic reduction of dimensionality in the electrodynamic problem for thin superconductors in transverse geometry [13], where the magnetic field has 3D distribution but the current density flow develops mainly parallel to the film plane. When the supercurrent distribution can be considered 2D, the $B_z$ component has logarithmic divergence at the sample edges because it results not-locally related to J(x,y); on the contrary, J(x,y) locally induces the in-plane magnetic field. Since, the $B_{xy}$ component reaches the maximum value of $\mu_0 J d/2$ at the superconductor surfaces, the in-plane fields saturate in the regions where the critical current developed, although in the same position the $B_z$ value can still increase. Therefore, when the external field is raised and more and more vortices nucleate into the superconductor, the magnetic field lines straighten from the edges toward the center.

The evolution of the magnetic field line curvature, as depending on vortex density and on position over the sample, can be best examined measuring the local variation of tilt angle $\theta$, as presented in figure 6. The $\theta(x,y_0)$ curve has non-linear monotonic increasing from edge toward the center in each quadrant of the squared superconducting film and it is a continuous function of the position, if the superconducting properties can be considered homogeneous. Apart from some defects labelled in the optical figure 1a), which perturb the Meissner current flow-path [16] and the subsequent local vortex arrangement, as visible in all the measurements, the superconducting properties of the sample can be assumed homogeneous on the length scale of our experiment.

Up to now, we dealt only with the curvature of the magnetic field lines, not the vortex one. In the "thin film" limit, i.e. when $d < \lambda$, where $\lambda$ is the London penetration depth, the vortices do not follow the magnetic field lines [4], because they have to cut the superconductor surface at right angles, whereas the magnetic field lines do not. For our films, with $d \geq 2\lambda$, the vortices may not be completely aligned with the magnetic field lines, but the lens-shaped 3D profile during vortex diffusion is fully developed. The in-plane magnetic field generate strong out-of-plane components of Lorenz force which change value and sign across the sample thickness as the in-plane fields do. Thus, anisotropic pinning sites may show up the influence of the magnetic field lines tilt on vortex pinning. We can check whether the pinned vortices are influenced by this magnetic field line tilt or not, throughout the study of local electrodynamic quantities, especially the local current density dependence by $\theta$.

## 4. Anisotropic pinning

The interplay between vortex curvature in transverse geometry and anisotropic pinning with respect to the *z* axis was theoretically treated in the work of Mikitik and Brandt [3]. They numerically solved the general electrodynamical problem of superconductors in transverse geometry, independently of the particular critical current dependence on $\theta$ and of the sample shape. It was pointed out the remarkable differences between pinning by defects correlated along the crystallographic *ab*-planes, such as intrinsic not superconducting planes of high temperature superconductors, and by defects extended along the *c* axis. In every case, the local maximum pinning efficiency, i.e. the local maximum critical current integrated over the thickness, is reached when the magnetic field line is directed along the defect easy axis, namely, where the Lorenz force is perpendicular to that direction.

Thus, we intend to investigate the pinning anisotropy of superconducting films by comparing the as-grown film and the same sample after the introduction of defects with known anisotropy. At first, the $B_z(x_0,y)$ and $J_x(x_0,y)$ components for the SQ2 sample, at T=5 K and several applied fields, are reported in figure 7a) and figure 7 b), respectively. Besides the improvement in the pinning capabilities of the system, revealed by the increased current density and the enhanced Meissner zone width, there are important differences in the shape of both profiles. The gradient of $B_z$ curves inside the sample is more linear after irradiation and the current density local values exhibit stronger depletion toward the Meissner area in comparison to the virgin state (without columns). These features could be attributed both to the known anisotropy of the extrinsic defects and to the creep process, which also results in smeared $J(x_0,y)$ profiles toward the flux-free area [17]. Thus, the observation of local electrodymanic variables as a function of position makes the two causes of J decreasing very difficult to be distinguished. On the contrary, by studying the local current density as a function of θ, in the vortex penetrated part, we are able to directly visualize the J dependence on vortex curvature. In figure 8, the local tilt angles are reported for the SQ2 sample, across the square mid-line of two defect-free quadrants. As discussed above, since in the studied samples the superconducting properties are quite homogeneous, the θ curve is a continuous and monotonic function of position in each quadrant and we thus obtain the corresponding J(θ) curves, presented in figure 9.

First of all, the macroscopic difference between the curves is represented by the current density gain with decreasing θ. The J(θ) in the irradiated SQ2 monotonically reaches the 180% of its value at the Meissner zone when we move toward the sample edges where the magnetic field lines are less curved with respect to the z axis. In the virgin SQ2, the maximum current gain is about 15% and the J(θ) trend neither appear to be monotonic.

The natural scaling of the several J(θ) curves traced for each applied field is another important experimental result contained in figure 9. The J(θ) collected data for each applied field contain a different number of points because we retain only the J values corresponding to the penetrated part of the superconductor, as we established from the $B_z=0$ criterion. Since the J(θ) curves, belonging to different external fields and different position inside the sample, overlap, we verified the vortex pinning is actually influenced by the magnetic field lines tilt in superconductors. Moreover, the effect of homogeneously distributed and strongly anisotropic defects is distinguished from naturally grown pinning sites in such YBCO films, which do not possess correlated defects as efficient as the irradiated induced columns. We speculate the natural pinning sites are distributed in a wider range of sizes and shapes with respect to the irradiated case, where the columns cleanly dominate the vortex pinning behaviour.

## 5. Discussion

From our analysis, it results the picture of straight flux lines originally suggested by Bean for infinitely elongated superconductors in parallel magnetic fields [18] does not hold for specimen of finite thickness in transverse geometry, where the flux line curvature causes a modulated critical current inside the specimen due to the anisotropy of pinning sites. As discussed above, the maximum current density, i.e. the critical current $J_C$, develops when the maximum pinning efficiency is reached. In the case of z axis correlated defects, this condition is fulfil for minimum vortex curvature. Therefore, we traced the $J(B_z)$ curves, one of them in figure 10, and retain as the critical current value the saturation threshold of the current density which corresponds to the minimum magnetic field line curvature into the sample.

Since after the applied field was held constant for a well defined waiting time (3 s), we acquire a snapshot of the magnetic field distribution inside a finite time window (<<1 s), i.e. while the creep process of vortices is in progress, the critical current values so defined should be considered as 'dynamical averaged'. The critical current, in fact, continuously decreases in non-linear way, as recently observed by magneto-optics in YBCO films [19]. Moreover, the time scales characteristic of the creep process may be different between the two cases (irradiated and virgin SQ2), thus, for true absolute comparison, the study of thermal fluctuations should be accompanied by relaxation measurements.

With this in mind, we plotted in figure 11 the critical current values at different temperatures, before and after irradiation for the SQ2 sample.

The huge gain (up to 100%) at low temperature settles down with increasing thermal excitation, then it raises again after about 30 K, as depicted in the inset of figure 11.

We expect, in single-vortex pinning state for irradiated SQ2, the creep process takes place by thermally activated motion of vortices or by their quantum tunnelling between column pin and neighbour sites. The steepest decreasing of $J_C$ at low temperatures indeed indicates the thermally activated motion of single vortices maybe is the main source of flux creep for the irradiated sample, even at the lowest temperature of our measurements, as seen before in the $J(x_0,y)$ profiles of figure 7b). At higher temperatures, the critical current behaviour is determined by the morphology of the extrinsic defects. We note that irradiation induced columnar defects have an insulating core surrounded by an amorphous region, which actually results in pinning potential wells of more complex shape than that of an ideal insulating cylinder [20]. Thus, the effect of thermal excitations may be more complicated than that predicted for ideal columnar pins [21], where size and morphology of the defect are well defined.

On the other hand, the columns cohabit with intrinsic defects and increase the total number of pins, despite of what is the efficiency condition. Then, we expect pinning capabilities of the irradiated system always better than the virgin one. At higher temperature, the raising gain may be due to collective pinning of the vortices with the columnar forest. Now, it is quite desirable to extend the QMO analysis to higher temperatures and to study in detail the electrodynamics of the superconducting as a function of the waiting time before the acquisition and of the field ramp rates.

### 6. Conclusion

We presented the experimental proof that magnetic field lines are curved in transverse geometry. For the first time, at our best knowledge, this curvature was locally revealed and quantified by means of QMO analysis. The local current density versus the magnetic field line tilt angle was found as the more convenient way to:

  i). to demonstrate how local and parallel measurements displays the not-local behaviour of electrodynamics of the transverse geometry;
 ii). to reveal the vortex curvature in presence of anisotropic pinning centers;
iii). to recognize different kind of pinning anisotropy, as the pinning due to extrinsic columnar defects or intrinsic correlated defects;

Moreover, we showed how local and parallel measurements yields a quite natural and model independent way to define a measured critical current, in transverse geometry and in presence of anysotropic defects. In the case of the strongly anisotropic columnar defects, we found possible signatures of vortex pinning-phase transition.

### Acknowledgments

We acknowledge the financial support of the Istituto Nazionale di Fisica Nucleare (INFN) and the Ministero dell'Istruzione, dell'Università e della Ricerca (MIUR). The YBCO samples were kindly supplied by the joined team "Edison-Europa Metalli-IMEM/CNR" under a partially funded CNR project (L.95/95).

**Figure Captions**

Figure 1. Optical pictures of the samples SQ1, a), and SQ2, b). The arrows in a) point out the defects which perturb the supercurrent flow path into the superconductor.

Figure 2. HRTEM results on YBCO twin samples. a) electron diffraction pattern showing the polycrystalline nature of our YBCO films. The grains have random in-plane orientation. b) <100> cross section at the YBCO/$CeO_2$ interface. An intrinsic amorphous defect is centered into the image. c) [001] planar view of the YBCO surface with a columnar defect (white area) produced by the Au irradiation. It is worthy to note that the columns are surrounded by an amorphous region, which results in enlarged and inhomogeneous defected zone.

Figure 3. The $B_z(x,y)$ distribution and the $J(x,y)$ stream lines of SQ1 YBCO film at T=5 K and applied external field of $\mu_0 H_a$=34.7 mT are shown. The sample was zero field cooled, afterwards, at constant temperature, the external field was raised by 3 mT steps up to the indicator film saturation field (~140 mT). The reported distributions correspond to a particular step during this virgin magnetization process.

Figure 4 a) 3D vector plot of B(x,y) at the surface of SQ1 in the same state as reported in figure 3. b) 2D vector plot of the in-plane magnetic field distribution, , on the film surface. The vectors are sampled over the shown area on a grid of 29x28 nodes, the distance between neighbour vector is 72 μm. Some of the area outside the sample is enclosed in the figures to remark the superposition of the external applied field and of the self-field of the sample.

Figure 5 $B_z(x,y_0)$ and $|B_{xy}(x,y_0)| = \left|\sqrt{B_x^2(x,y_0) + B_y^2(x,y_0)}\right|$ values, 5a) and 5b) respectively, in function of position over the SQ1 sample at T = 5 K and different applied field. The profiles were traced over the mid line of the square, thus can be compared to those of thin superconductors in strip geometry.

Figure 6. $\theta(x,y_0)$ values at T = 5 K and several external applied fields, same position as before. The discontinuity pointed by the arrow indicates the effect of a defect on θ. The missing points in the central parts belong to the Meissner zone where there are no vortices ($B_z$=0). The not-zero curvature outside the sample is due to spurious in-plane magnetic domains of the indicator film.

Figure 7. $B_z(x_0,y)$ (a) and $J_x(x_0,y)$ (b) profiles for irradiated and virgin SQ2 sample at T=5 K and several applied fields. As before, the profiles were traced along the mid line of the square. Some differences between the effect due to curved vortices interacting with *z* axis correlated pinning centers results and the vortex creep could be distinguished going toward the Meissner area, where the $J_x(x_0,y)$ trend in is steeper for curvature effects than the creep caused ones and it should be present a 'characteristic knee' in $J(x_0,y)$ curve near the Meissner zone boundary, see the right part of figure 7b).

Figure 8. $\theta(x,y_0)$ values at T = 5 K and several external applied fields for virgin and irradiated SQ2.

Figure 9. $J(\theta)$ curves at T = 5 K for virgin and irradiated SQ2. Note that several curves overlap, independently of the applied field which belong to. Also the number of point for each applied field is different, thus the $J(\theta)$ curve are naturally scaling.

Figure 10. J(Bz) and J(Bz(x0,y)) profile at T = 5K and $\mu_0H_a$=72 mT for $J(B_z(x_0,y))$. The $J(B_z)$ curve is taken in a specified point inside the sample, i.e. it is the local J value as a function of the local Bz which changes in correspondence of the applied magnetic field, whereas the $J(B_z(x_0,y))$ curve is the data set got from a single image, i.e. the local J as a function of the local $B_z$ which changes on position. Because of the homogeneity of the superconducting properties, the two curves overlaps, just demonstrating how our QMO method correctly evaluates the actual electrodynamics quantities in each point inside the sample.

Figure 11. $J_C(T)$ for irradiated and virgin SQ2. The values are measured as specified into the text. Inset, linear gain between the irradiated and virgin sample as a function of temperature.

**Figures**

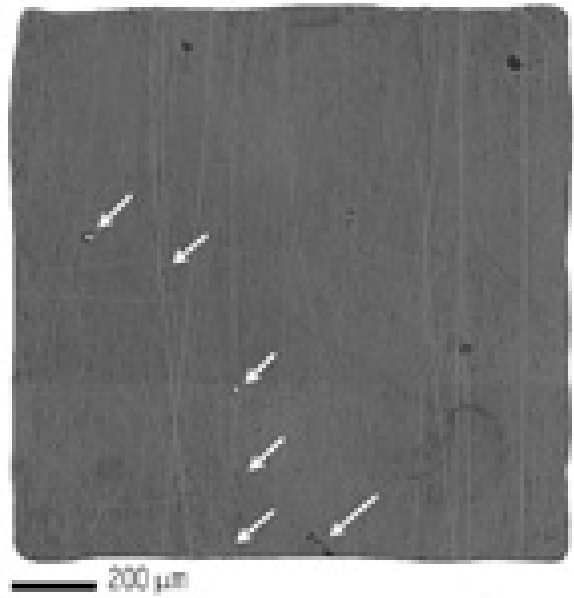

a)

Figure 1.

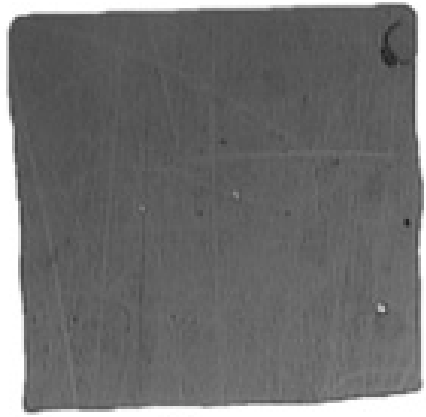 b)

Figure 1.

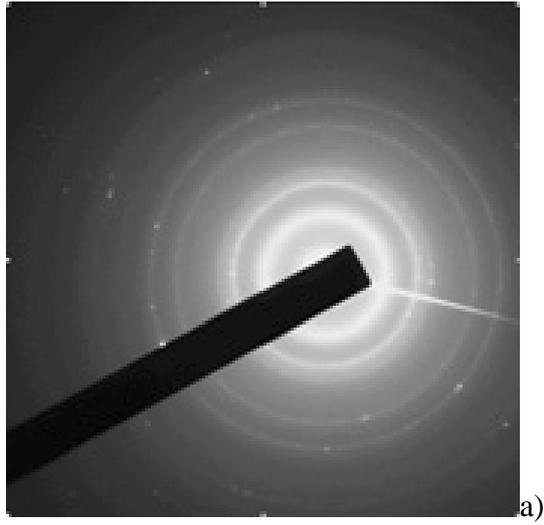a)

Figure2.

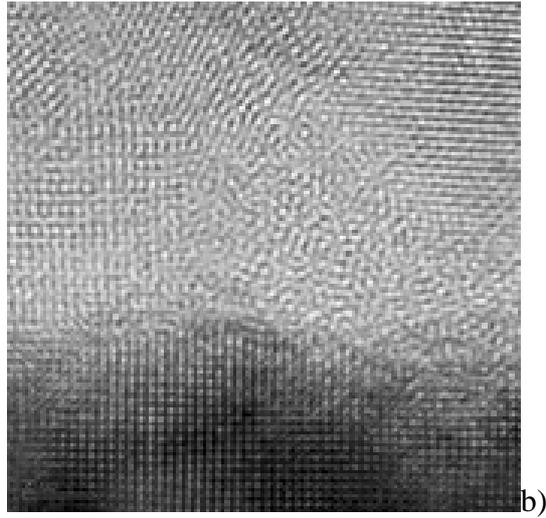

Figure2.

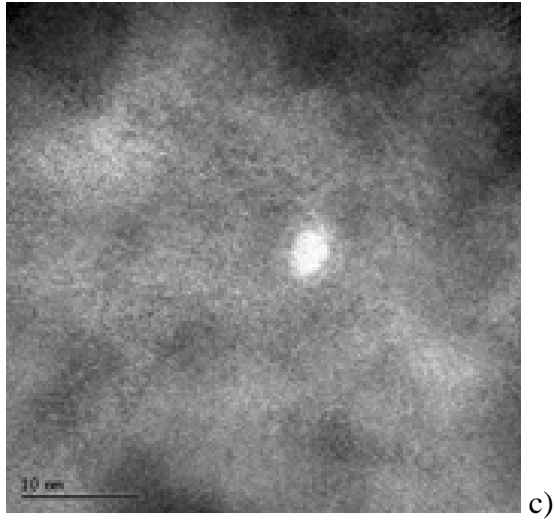

Figure 2.

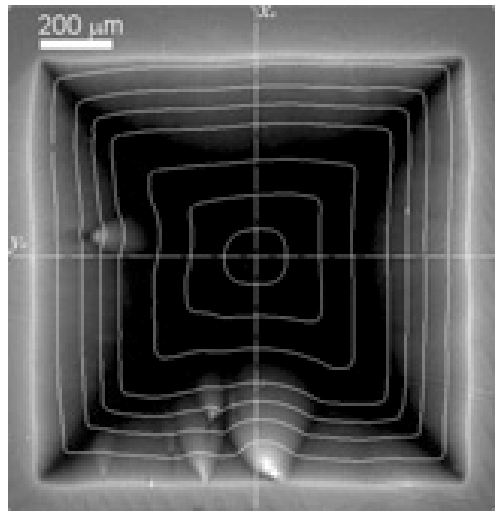

Figure 3.

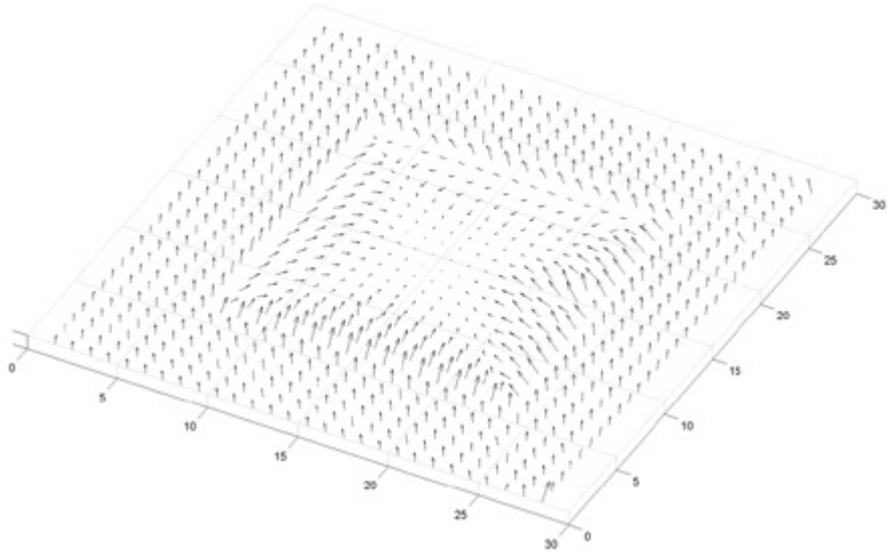

a)

Figure 4

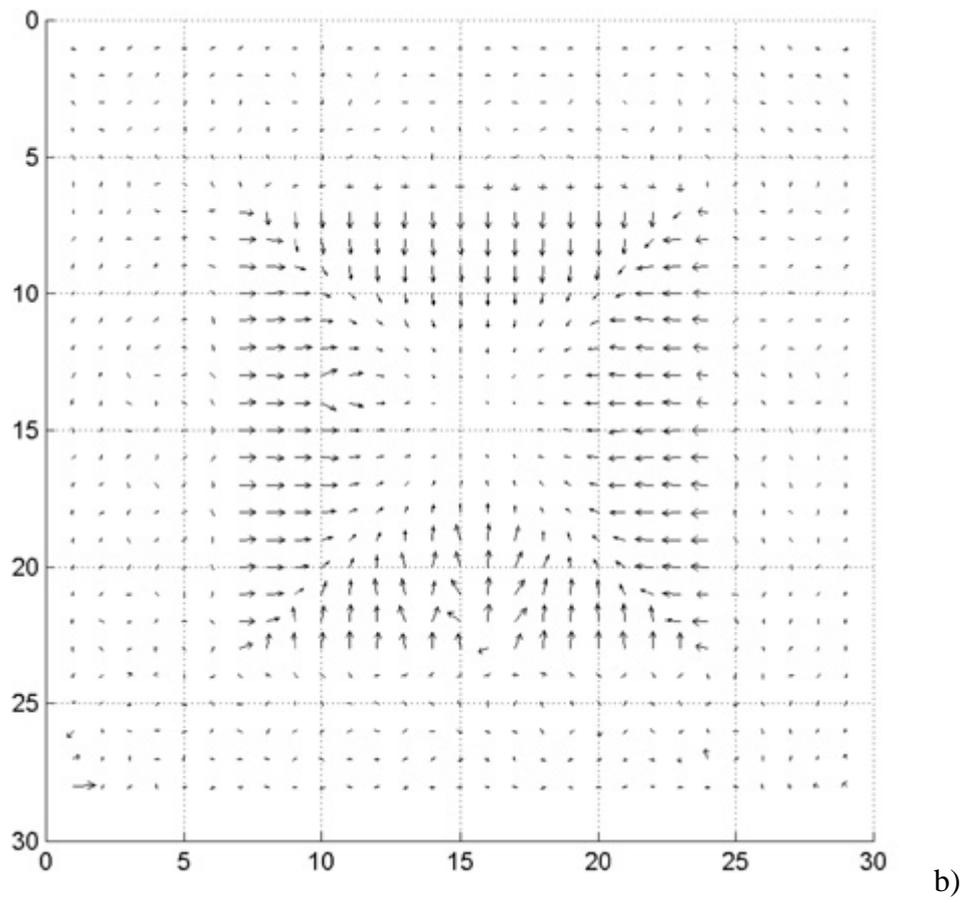

b)

Figure 4.

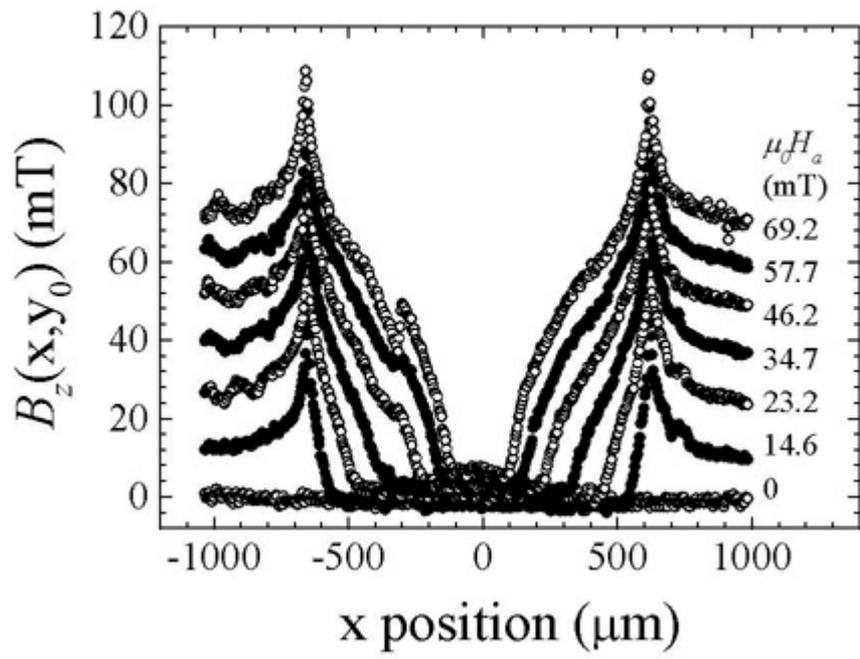

a)

Figure 5.

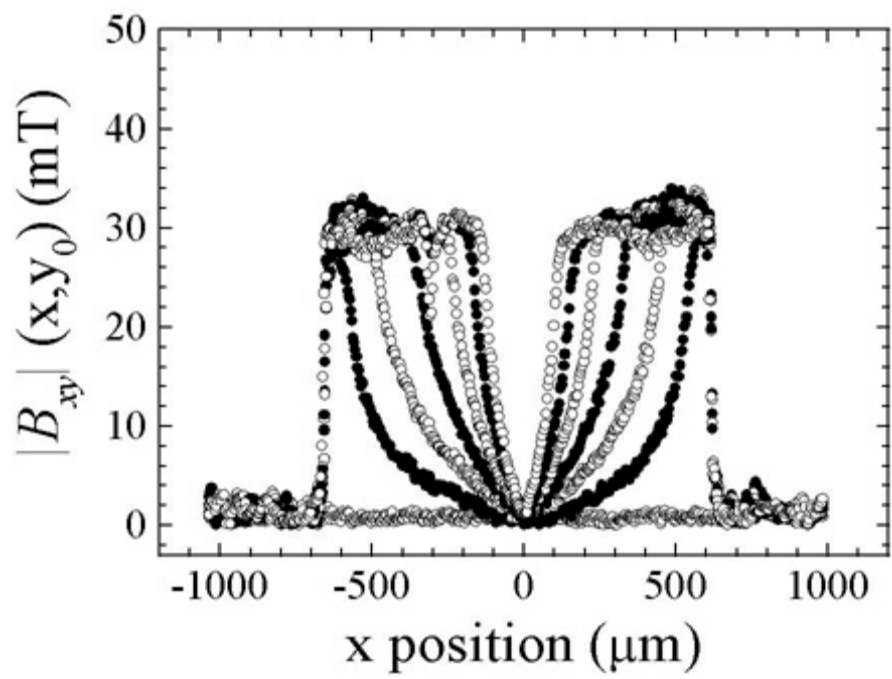 b)

Figure 5.

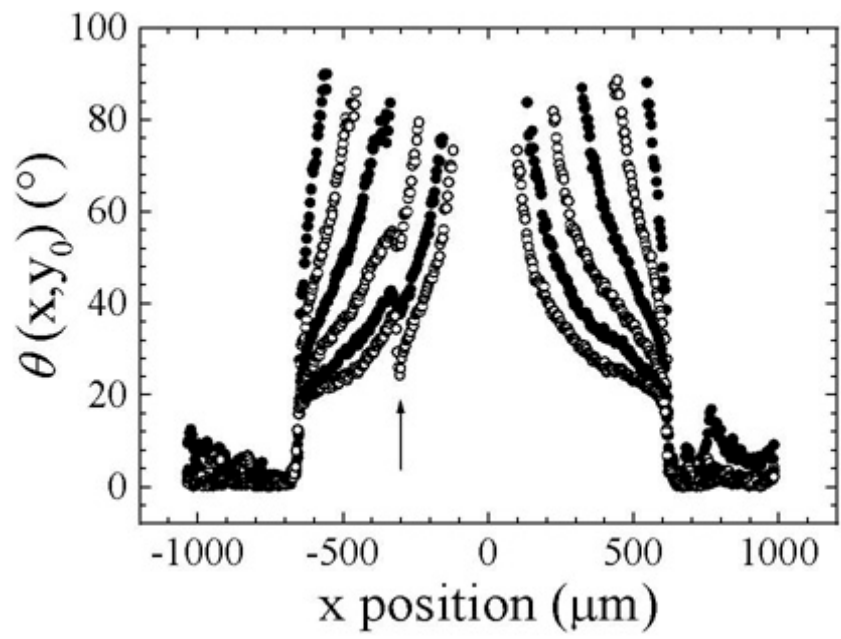

Figure 6.

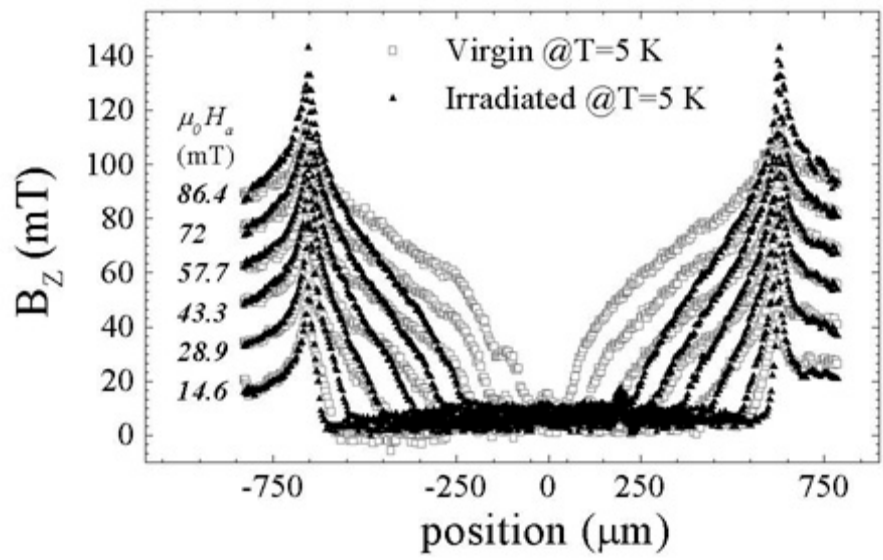

a)

Figure 7.

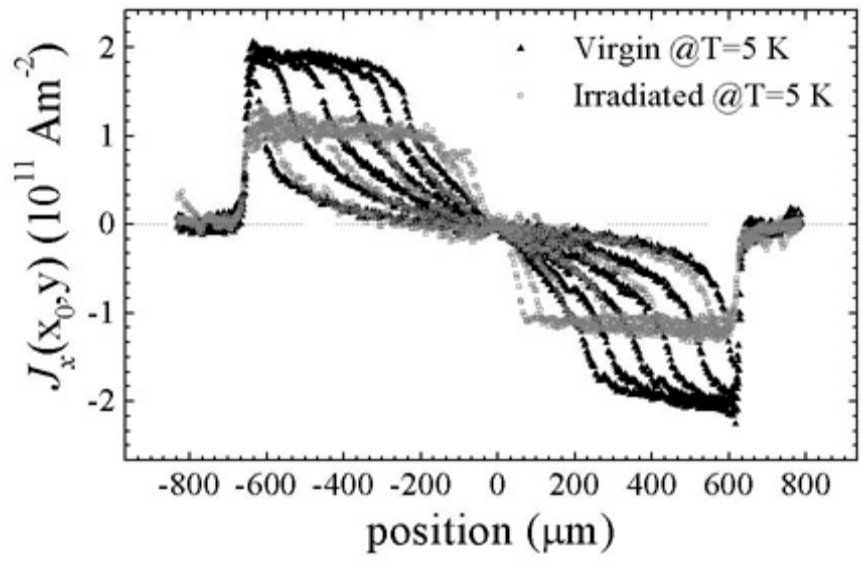

b)

Figure 7.

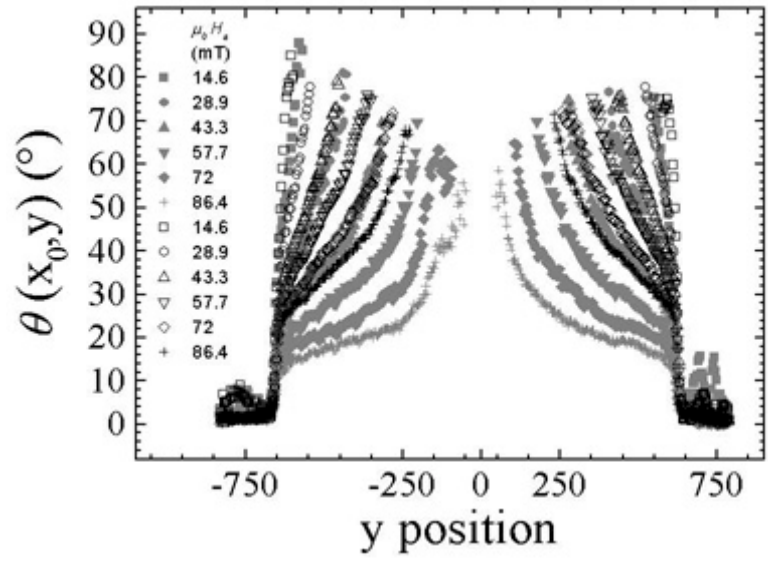

Figure 8.

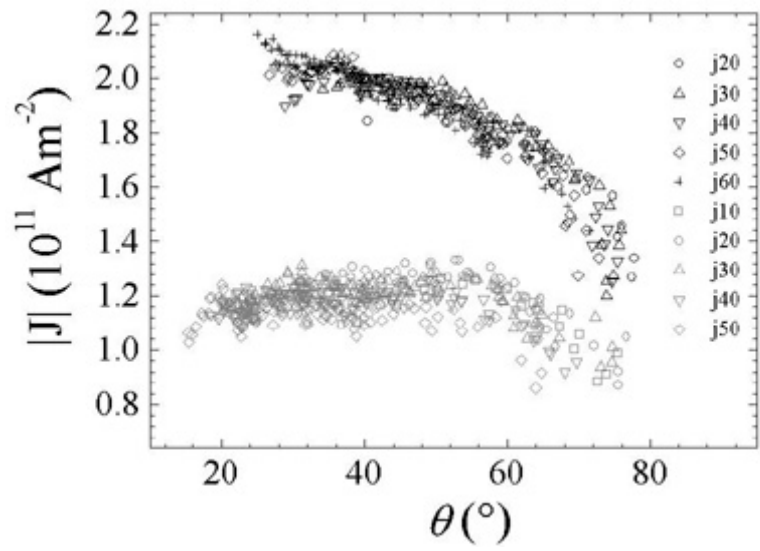

Figure 9.

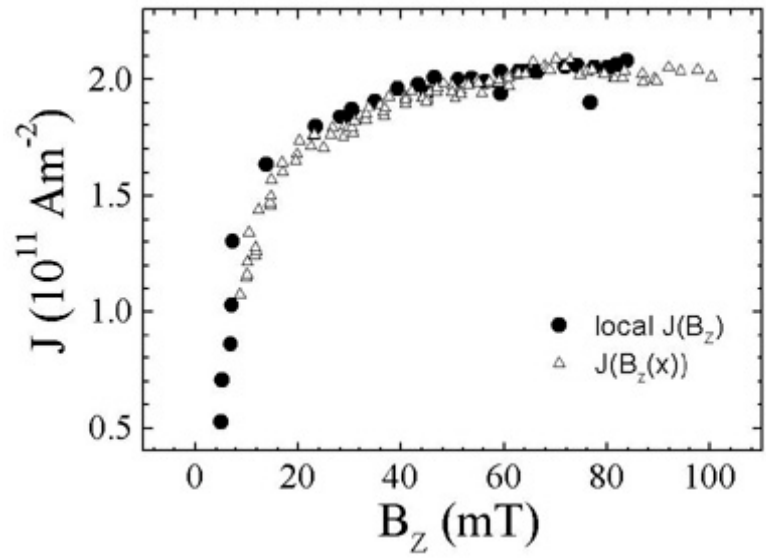

Figure 10.

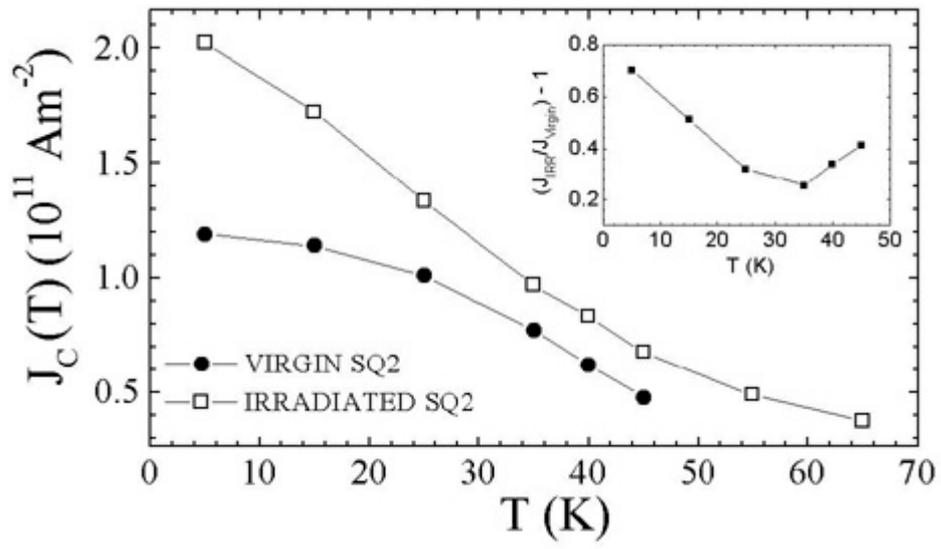

Figure 11.